\documentclass{cimento}%

\usepackage{aas_macros}
\usepackage{amsfonts}
\usepackage{amsmath}
\usepackage{amssymb}
\usepackage{graphicx}

\title{Electron-positron plasma in GRBs and in cosmology}

\author{R.~Ruffini\from{ins:x}\from{ins:y} \atque
G.~V.~Vereshchagin\from{ins:x}\from{ins:y}}
\instlist{
\inst{ins:x} Dipartimento di Fisica, Sapienza Universit\`a di Roma, ICRA - International Center for Relativistic Astrophysics, P.le Aldo Moro 5, Rome, 00185, Italy
\inst{ins:y} ICRANET - P.zza della Repubblica 10, Pescara, 65122, Italy}

\PACSes{ 
\PACSit{52.27.Ep}{Electron-positron plasmas}
\PACSit{52.27.Ny}{Relativistic plasmas}
\PACSit{98.80.-k}{Cosmology}}
\begin{document}

\maketitle

\begin{abstract}
Electron-positron plasma is believed to play imporant role both in the early
Universe and in sources of Gamma-Ray Bursts (GRBs).
We focus on analogy and difference between physical conditions of
electron-positron plasma in the early Universe and in sources of GRBs.
We discuss a) dynamical differences, namely thermal acceleration of the
outflow in GRB sources vs cosmological deceleration; b) nuclear composition
differences as synthesis of light elements in the early Universe and possible
destruction of heavy elements in GRB plasma; c) different physical conditions
during last scattering of photons by electrons. Only during the acceleration
phase of the optically thick electron-positron plasma comoving observer may
find it similar to the early Universe. This similarity breaks down during the
coasting phase. Reprocessing of nuclear abundances may likely take place in
GRB sources. Heavy nuclear elements are then destroyed, resulting mainly in
protons with small admixture of helium. Unlike the primordial plasma which
recombines to form neutral hydrogen, and emits the Cosmic Microwave Background
Radiation, GRB plasma does not cool down enough to recombine.
\end{abstract}

\section{Introduction}

Electron-positron plasmas are discussed in connection with astrophysical
phenomena such as Galactic Center, microquasars, Gamma-Ray Bursts (GRBs), as
well as laboratory experiments with high power lasers, for details see
\cite{Ruffini2009}. According to the standard cosmological model, such plasma
existed also in the early Universe. It is naturally characterized by the
energy scale given by the electron rest mass energy, $511$ keV. It is
interesting that at the epoch when Universe had this temperature, several
important phenomena took place almost contemporarily: electron-positron pair
annihilation, the Big Bang Nucleosynthesis (BBN) and neutrino decoupling.

Electron-positron plasma also is thought to play an essential role in GRB
sources, where simple estimates for the initial temperature give values in MeV
region. Such plasma is energy dominated and optically thick due to both
Compton scattering and electron-positron pair creation, and relaxes to thermal
equilibrium on a time scale less than $10^{-11}$ sec, see
\cite{2007PhRvL..99l5003A}. The latter condition results in self-accelerated
expansion of the plasma until it becomes either transparent or matter dominated.

In the literature there have been several qualitative arguments mentioning
possible similarities between electron-positron plasmas in the early Universe
and in GRB sources. However, until now there is no dedicated study which draws
analogies and differences between these two cases. This paper aims in
confronting dynamics and physical conditions in both cases.

\section{General equations}

The framework which describes electron-positron plasma both in cosmology and
in GRB\ sources is General Relativity. Both dynamics of expansion of the
Universe, and the process of energy release in the source of GRB should be
considered within that framework. Hydrodynamic expansion of GRB sources may,
however, be studied within much simplier formalism of Special Relativity.

We start with Einstein equations%
\begin{equation}
R_{\mu\nu}-\frac{1}{2}g_{\mu\nu}R=\frac{8\pi G}{c^{4}}T_{\mu\nu}, \label{EE}%
\end{equation}
where $R_{\mu\nu}$, $g_{\mu\nu}$\ and $T_{\mu\nu}$\ are respectively Ricci,
metric and energy-momentum tensors, $G$ is Newton's constant, $c$ is the speed
of light, and the energy-momentum conservation, following from (\ref{EE})%
\begin{equation}
\left(  T_{\mu}~^{\nu}\right)  _{;\nu}=\frac{1}{\sqrt{-g}}\,\frac{\partial(\sqrt{-g}\,T_{\mu}%
~^{\nu}{})}{\partial x^{\nu}}-\Gamma_{\nu\mu}^{\lambda}%
T_{\lambda}{}^{\nu}=0, \label{EM}%
\end{equation}
where $\Gamma_{\nu\lambda}^{\mu}$\ are Cristoffel symbols and $g$ is
determinant of the metric tensor. We assume for the energy-momentum tensor%
\begin{equation}
T^{\mu\nu}=p\,g^{\mu\nu}+\omega U^{\mu}U^{\nu},
\end{equation}
where $U^{\mu}$, is four-velocity, $\omega=\rho+p$ is proper enthalpy, $p$ is
proper pressure and $\rho$ is proper energy density.

When plasma is optically thick, radiation is trapped in it and entropy
conservation applies. It may be obtained multiplying (\ref{EM}) by
four-velocity%
\begin{equation}
-U^{\mu}\left(  T_{\mu}~^{\nu}\right)  {}_{;\nu}=U^{\mu}\rho{}_{;\mu}+\omega
U^{\mu}{}_{;\mu}=0. \label{entrcons}%
\end{equation}
Using the second law of thermodynamics%
\begin{equation}
d\left(  \frac{\omega}{n}\right)  =Td\left(  \frac{\sigma}{n}\right)
+\frac{1}{n}dp,
\end{equation}
where $\sigma=\omega/T$ is proper entropy density, $T$ is temperature, one may
rewrite (\ref{entrcons}) as%
\begin{equation}
(\sigma U^{\mu})_{;\mu}=U^{\mu}\sigma{}_{;\mu}+\sigma U^{\mu}{}_{;\mu}=0.
\label{sigmacons}%
\end{equation}

Baryon number conservation equation has exactly the same form%
\begin{equation}
(nU^{\mu})_{;\mu}=U^{\mu}n{}_{;\mu}+nU^{\mu}{}_{;\mu}=0. \label{numcons}%
\end{equation}

Now recalling that $U^{\mu}\frac{\partial}{\partial x^{\mu}}=\frac{d}{d\tau}$
and $U^{\mu}{}_{;\mu}=d\ln V/d\tau$, where $V$\ is comoving volume, $\tau$ is
the proper time, from (\ref{entrcons}) and (\ref{numcons}) we get%

\begin{equation}
d\rho+\omega d\ln V=0,\quad\quad d\ln n+d\ln V=0,
\end{equation}
Finally, introducing the thermal index $\gamma=1+\frac{p}{\rho}$ restricted by
the inequality $1\leq\gamma\leq4/3$ we obtain from\ (\ref{entrcons}) the
following scaling laws%
\begin{equation}
\rho V^{\gamma}=\mathrm{const},\quad\quad nV=\mathrm{const}. \label{num}%
\end{equation}
Both these conservations laws are valid for the early Universe and GRB plasmas.

One can obtain the corresponding scaling laws for comoving temperature by
splitting the total energy density into nonrelativistic (with $\gamma=1$) and
ultrarelativistic (with $\gamma=4/3$) parts with $\rho\rightarrow nmc^{2}%
+\varepsilon$, where $m$\ is the mass of particles\footnote{Nonrelativistic
component is represented by baryons. For simplicity we assume only one sort of
baryons, say protons, having mass $m$. Ultrarelativistic component is
represented by photons and electron-positron pairs.}, $\varepsilon$ is proper
internal energy density. The entropy of the ultrarelativistic component is
then $\sigma=\frac{4}{3}\frac{\varepsilon}{T}$, and (\ref{sigmacons}) gives%
\begin{equation}
\frac{\varepsilon V}{T}=\mathrm{const}. \label{temp}%
\end{equation}
For $\varepsilon\gg nmc^{2}$, which is the energy dominance condition,
internal energy plays dynamical role by influencing the laws of expansion. For
$\varepsilon\ll nmc^{2}$, which is the matter dominance condition, internal
energy does not play any dynamical role, but determines the scaling law of the
temperature. In order to understand the dynamics of thermodynamic quantities
in both early Universe and in GRBs, one should write down the corresponding
equations of motion.

\subsection{Early Universe}

For the description of the early Universe we take the Robertson-Walker metric
with the interval%
\begin{equation}
ds^{2}=-c^{2}dt^{2}+a^{2}\left(  t\right)  \left[  \frac{dr^{2}}{1-kr^{2}%
}+r^{2}d\vartheta^{2}+r^{2}\sin^{2}\vartheta d\varphi^{2}\right]  ,
\label{RWcoord}%
\end{equation}
where $a\left(  t\right)  $ is the scale factor and $k=0,\pm1$\ stands for the
spatial curvature.\ In homogeneous and isotropic space described by
(\ref{RWcoord}), Einstein equations (\ref{EE})\ are reduced to Firedmann
equations together with the continuity equation%
\begin{align}
\left(  \frac{da}{dt}\right)  ^{2}+c^{2}k  &  =\frac{8\pi G}{3c^{2}}\rho
a^{2},\label{FE}\\
2a\frac{d^{2}a}{dt^{2}}+\left(  \frac{da}{dt}\right)  ^{2}+c^{2}k  &
=-\frac{8\pi G}{c^{2}}pa^{2},\label{FE2}\\
\frac{d\rho}{dt}+\frac{3}{a}\frac{da}{dt}\left(  \rho+p\right)   &  =0,
\label{cosmcont}%
\end{align}
where $a$\ is the scale factor. Notice, that only two equations in the system
above are independent. The continuity equation (\ref{cosmcont}) follows from
the Einstein equations (\ref{FE}) and (\ref{FE2}) as the energy conservation.
In fact, (\ref{cosmcont}) may be also obtained from the entropy conservation
(\ref{entrcons}). The comoving volume in Friedmann's Universe scales with $a$ as $V=a^{3}$, so
(\ref{cosmcont}) and the first equality in (\ref{num}) are equivalent.

On the radiation dominated stage of the Universe expansion one has%
\begin{equation}
\rho\propto V^{-4/3}\propto a^{-4},\quad\quad n\propto V^{-1}\propto a^{-3},
\label{enU}%
\end{equation}
while on the matter dominated stage%
\begin{equation}
\rho\propto n\propto V^{-1}\propto a^{-3}. \label{ennumU}%
\end{equation}
Entropy conservation (\ref{temp}) leads to the unique temperature dependence
on the scale factor%
\begin{equation}
T\propto V^{-1/3}\propto a^{-1}. \label{tempU}%
\end{equation}

The corresponding time dependence of thermodynamic quantities may be obtained
from solutions of Friedmann equation (\ref{FE}) and continuity equation
(\ref{cosmcont}), see e.g. \cite{2008gcpa.book.....W}.

\subsection{GRBs}

Different situation takes place for the sources of GRBs. Assuming spherical
symmetry for the case of GRB the interval\footnote{General Relativity effects
may be included by taking Schwarzschild or Kerr-Newman metric. However, we are
interested in optically thick plasma which expands with acceleration and
propagates far from its source, where the spatial curvature effects may be
neglected. For this reason we simplify the treatment and adopt a spatially
flat metric.} is
\begin{equation}
ds^{2}=-c^{2}dt^{2}+dr^{2}+r^{2}d\vartheta^{2}+r^{2}\sin^{2}\vartheta
d\varphi^{2}. \label{sphericalcoord}%
\end{equation}
Optically thick to Compton scattering and pair production electron-positron
plasma in GRB sources is radiation dominated. Its equations of motion follow
from the energy-momentum conservation law (\ref{EM}) and baryon number
conservation law (\ref{numcons}). Initially plasma expands with acceleration
driven by the radiative pressure.

In spherically symmetric case the number conservation equation (\ref{numcons}) is%

\begin{equation}
\frac{\partial\left(  n\Gamma\right)  }{\partial t}+\frac{1}{r^{2}}%
\frac{\partial}{\partial r}\left(  r^{2}n\sqrt{\Gamma^{2}-1}\right)  =0,
\label{continuityeq}%
\end{equation}
Integrating this equation over the volume from certain $r_{i}(t)$ to
$r_{e}(t)$ which we assume to be comoving with the fluid $\frac{dr_{i}(t)}%
{dt}=\beta(r_{i},t)$, $\frac{dr_{e}(t)}{dt}=\beta(r_{e},t)$, and ignoring a
factor $4\pi$ we have%

\begin{gather}%
{\displaystyle\int\limits_{r_{i}}^{r_{e}}}
\frac{\partial\left(  n\Gamma\right)  }{\partial t}r^{2}dr+%
{\displaystyle\int\limits_{r_{i}}^{r_{e}}}
\frac{\partial}{\partial r}\left(  r^{2}n\sqrt{\Gamma^{2}-1}\right)
dr=\label{nslabcon}\\
\frac{\partial}{\partial t}%
{\displaystyle\int\limits_{r_{i}}^{r_{e}}}
\left(  n\Gamma\right)  r^{2}dr-\frac{dr_{e}}{dt}n(r_{e},t)\Gamma
(r_{e},t)r_{e}^{2}+\frac{dr_{i}}{dt}n(r_{i},t)\Gamma(r_{i},t)r_{i}%
^{2}+\nonumber\\
+r_{e}^{2}n(r_{e},t)\sqrt{\Gamma^{2}(r_{e},t)-1}-r_{i}^{2}n(r_{i}%
,t)\sqrt{\Gamma^{2}(r_{i},t)-1}=\nonumber\\
=\frac{d}{dt}%
{\displaystyle\int\limits_{r_{i}}^{r_{e}}}
\left(  n\Gamma\right)  r^{2}dr=0,\nonumber
\end{gather}
Since we deal with arbitrary comoving boundaries, this means that the total
number of particles integrated over all differential shells is conserved%

\begin{equation}
N=4\pi%
{\displaystyle\int\limits_{0}^{R(t)}}
n\Gamma r^{2}dr=\mathrm{const}, \label{Ncons}%
\end{equation}
where $R(t)$ is the external radius of the shell.

Following \cite{1993MNRAS.263..861P} one can transform (\ref{continuityeq})
from the variables $(t,r)$ to the new variables $(s=t-r,r)$ and then show that%
\begin{equation}
\frac{1}{r^{2}}\frac{\partial}{\partial r}\left(  r^{2}n\sqrt{\Gamma^{2}%
-1}\right)  =-\frac{\partial}{\partial s}\left(  \frac{n}{\Gamma+\sqrt
{\Gamma^{2}-1}}\right)  . \label{transform}%
\end{equation}
For ultrarelativistic expansion velocity $\Gamma\gg1$, the RHS in
(\ref{transform})\ tends to zero, and then the number of particles in each
differential shell between the boundaries $r_{i}(t)$ and $r_{e}(t)$ is also
conserved with a good approximation, i.e.%

\begin{equation}
dN=4\pi n\Gamma r^{2}dr\approx\mathrm{const}. \label{dNcons}%
\end{equation}
Relations (\ref{Ncons}) and (\ref{dNcons}) then imply%

\begin{align}
4\pi%
{\displaystyle\int\limits_{r_{i}}^{r_{e}}}
\left(  n\Gamma r^{2}\right)  dr  &  =4\pi\left[  n(r,t)\Gamma(r,t)r^{2}%
\right]
{\displaystyle\int\limits_{r_{i}}^{r_{e}}}
dr\\
&  =4\pi\left(  n\Gamma r^{2}\right)  \Delta\approx\mathrm{const},\nonumber
\end{align}
where the first argument of functions $n(r,t)$ and $\Gamma(r,t)$ is restricted
to the interval $r_{i}<r<r_{e}$, and consequently $\Delta\equiv r_{e}%
-r_{i}\approx\mathrm{const}$. Taking into account that $r_{i}(t)$ and
$r_{e}(t)$ are arbitrary, this means that ultrarelativistically expanding
shell preserves its width measured in the laboratory reference frame. This
fact has been used in \cite{2000A&A...359..855R} and referred there as the
constant thickness approximation.

The volume element measured in the laboratory reference frame is
$d\mathcal{V}=4\pi r^{2}dr$, while the volume element measured in the
reference frame comoving with the shell is $dV=4\pi\Gamma r^{2}dr$. Comoving
volume of the expanding ultrarelativistic shell with $\Gamma\simeq
\mathrm{const}$\ will be
\begin{equation}
V=4\pi\Gamma\int_{r-\Delta}^{r}r^{2}dr\simeq4\pi\Gamma r^{2}\Delta.
\end{equation}
Then we rewrite the conservation equations (\ref{num}) as%
\begin{equation}
\rho^{\frac{1}{\gamma}}\Gamma r^{2}=\mathrm{const},\quad\quad n\Gamma
r^{2}=\mathrm{const}, \label{entrGRB}%
\end{equation}
Unlike the early Universe, where both energy and entropy conservations reduce
to (\ref{cosmcont}), in the case of GRBs the energy conservation is a separate
equation coming from the zeroth component of (\ref{EM}) as%
\begin{equation}
\left(  T_{0}~^{\nu}\right)  {}_{;\nu}=\omega U_{0}U^{\nu}{}_{;\nu}+U^{\nu
}\left(  \omega U_{0}\right)  _{;\nu}=0. \label{encons}%
\end{equation}
which independently of $\gamma$ gives%
\begin{equation}
\rho\Gamma^{2}r^{2}=\mathrm{const}. \label{enGRB}%
\end{equation}
From (\ref{entrGRB}) and (\ref{enGRB}) we then find%
\begin{equation}
\Gamma\propto r^{\frac{2\left(  \gamma-1\right)  }{2-\gamma}},\quad n\propto
r^{-\frac{2}{2-\gamma}},\quad\rho\propto r^{-\frac{2\gamma}{2-\gamma}}.
\end{equation}
For the ultrarelativistic equation of state with $\gamma=4/3$ we immediately
obtain%
\begin{equation}
\Gamma\propto r,\quad n\propto r^{-3},\quad\rho\propto r^{-4}.
\label{scalingsUR}%
\end{equation}
Taking into account that the relation between the comoving and the physical
coordinates in cosmology is given by the scale factor $a$,\ it follows from
(\ref{scalingsUR}) that both energy density and baryonic number density behave
as in the radiation dominated Universe, see (\ref{enU}). This analogy between
the GRB source and the Friedmann Universe is noticed by
\cite{1990ApJ...365L..55S}, \cite{1993MNRAS.263..861P}.

In the presence of baryons as the pressure decreases, plasma becomes matter
dominated and expansion velocity saturates. Hence for the nonrelativistic
equation of state with $\gamma=1$ different scaling laws come out%
\begin{equation}
\Gamma=\mathrm{const},\quad n\propto r^{-2},\quad\rho\propto r^{-2}.
\label{scalingsNR}%
\end{equation}
Transition between the two regimes (\ref{scalingsUR}) and (\ref{scalingsNR})
occurs at the radius $R_{c}=B^{-1}R_{0}$, where $R_{0}$\ is initial size of plasma.

Therefore, one may reach the conclusion that for comoving observer the
radiation-dominated plasma looks indistinguishable from a portion of
radiation-dominated Universe. However, this is true only in the absence of
pressure gradients. Strong gradients are likely present in GRB sources, and
they should produce local acceleration in the radiation-dominated
electron-positron plasma, making it distinct from the early Universe, where
matter inhomogeneities are known to be weak.

It is easy to get from (\ref{entrGRB}) and (\ref{enGRB}) for internal energy
density and temperature%

\begin{equation}
\varepsilon\propto r^{-4},\quad T\propto r^{-1},\quad R_{0}<r<R_{c},
\label{acceleration}%
\end{equation}
and
\begin{equation}
\varepsilon\propto r^{-8/3},\quad T\propto r^{-2/3},\quad R_{c}<r<R_{tr},
\label{coasting}%
\end{equation}
where $R_{tr}$ is the radius at which the outflow becomes transparent. The
outflow may become transparent for photons also at the acceleration phase,
provided that $R_{tr}<R_{c}$. For instance, a pure electron-positron plasma
gets transparent at the acceleration phase.

\section{Heavy elements}

Cosmological nucleosynthesis is a well established branch of cosmology.
Classical computations made in the middle of the XXth century revealed that
heavy elements cannot be built in the early Universe. Hydrogen and helium
contribute approximately 3/4 and 1/4, leaving some room, much less than 1 per
cent for deuterium, tritium and lithium. All the heavier elements must have
been produced in stars.

Some of these stars, as indicated by observations, end their life as
progenitors of GRBs. For this reason it is likely that initially in the source
of GRBs elements heavier than hydrogen are present. In this section we
consider chemical evolution of plasma in the sources of GRBs.

Assume that in the source of a GRB the amount of energy $E_{0}$ is released in
the volume with linear size $R_{0}$ during the time $\Delta t$, making this
region optically thick to Compton scattering and pair production. The amount
of baryons which may be present as well is parametrized by%

\begin{equation}
B\simeq\left\{
\begin{array}
[c]{cc}%
\dfrac{Mc^{2}}{E_{0}}, & \Delta t\ll R_{0}/c,\\
& \\
\dfrac{\dot{M}c^{2}}{L}, & \Delta t\gg R_{0}/c,
\end{array}
\right.  \label{Bdef}%
\end{equation}
where $L=dE/dt$ is the luminosity, $\dot{M}=dM/dt$ is the mass ejection rate
and $M$\ is total baryonic mass. Ultrarelativistic outflow is generated
through thermal acceleration of baryons by the radiative pressure if plasma is
initially energy dominated, i.e.%
\begin{equation}
B\ll1. \label{Bconstraint}%
\end{equation}
In the case of instant energy release with time interval $\Delta t\sim R_{0}/c$\ initial
temperature in the source of GRB may be estimated neglecting the baryonic
contribution, provided (\ref{Bconstraint}) is satisfied as%
\begin{equation}
T_{0}\simeq\left(  \dfrac{3E_{0}}{4\pi aR_{0}^{3}}\right)  ^{1/4}%
\simeq6.5E_{54}^{1/4}R_{8}^{-3/4}\;\mathrm{MeV}, \label{T0}%
\end{equation}
where $a=4\sigma_{SB}/c$, $\sigma_{SB}$\ is the Stefan-Boltzmann constant and
the last value is obtained by substituting numerical values for $E_{0}%
=10^{54}E_{54}$ erg and $R_{0}=10^{8}R_{8}$ cm.

As it has been shown in \cite{2010tsra.confE.274K} for temperatures above 1
MeV even low density plasma with density $n=10^{18}$\ cm$^{-3}$ quite quickly
destroys all heavier nuclei, and the final state contains just protons and
neutrons and some small traces of Deuterium and $^{4}$He. The timescale of
this process ($\sim10^{-2}$ sec for $T_{0}=1$ MeV) strongly depends on
temperature, but the rates of almost all reactions increase with temperature,
and correspondingly the abundances of nuclei evolve much faster. Therefore,
nuclei disintegration is fast enough to occur before plasma starts to expand
and cool on the timescale $R_{0}/c$.

During early stages of plasma expansion its temperature decreases in the same
way as it happened in the early Universe. Therefore similar synthesis of light
elements to BBN\ occurs also in sources of GRBs. Most important is, however,
another similarity with the early Universe: it is well known that practically
all free neutrons have been captured into elements heavier than hydrogen. So
we do not expect dynamically important free neutrons present in GRB plasma
after it started to expand and cool down unless they are engulfed by the
expanding plasma later. The role of such free neutrons have been considered in
the literature, see e.g. \cite{2000NuPhS..80C0612D} and \cite{2010MNRAS.407.1033B}.

\section{Recombination}

On the radiation dominated phase both in the early Universe and in GRB plasma
entropy conservation (\ref{entrcons}) results in decrease of temperature. When the comoving temperature decreases below the hydrogen
ionization energy, $E_{i}=13.6$\ eV, the formation of neutral hydrogen is expected.

\subsection{Early Universe}

In the early Universe, after the BBN epoch and electron-positron annihilation,
cosmological plasma consists of fully ionized hydrogen, helium and small
admixture of other light elements. The temperature continues to decrease until
it gets sufficiently low to allow formation of neutral atoms:\ that is the
moment in the cosmic history where the formation of the Cosmic Microwave
Background Radiation (CMB) happens.

The theory of cosmological recombination of hydrogen, based on three level
approximation, has been developed in \cite{1968ZhETF..55..278Z} and
\cite{1968ApJ...153....1P}\ in the late 60s. The only modification that such
theory undergone in the later years is the account for dark matter and
addition of more levels to the model, currently about 300. There is a basic
difference with respect to the equilibrium recombination essentially by the
process $e+p\leftrightarrow H+\gamma$, described by the Saha equation%
\begin{equation}
\frac{n_{e}n_{p}}{n_{H}}=\frac{g_{e}g_{p}}{g_{H}}\frac{\left(  2\pi
m_{e}kT\right)  ^{3/2}}{h^{3}}\exp\left(  -\frac{E_{i}}{kT}\right)  ,
\end{equation}
where $g_{i}$\ are statistical weights, $h$\ is Planck's constant. This
difference is due to the presence of the $2p$ quantum level, which produces
Ly-$\alpha$ photons. The absorption of such photons is very strong. However,
ionization from the $2p$\ level requires only $E_{i}/4$. Therefore the
formation of neutral hydrogen proceeds through the $2s-1s$ transition in the
presence of abundant Ly-$\alpha$ photons.

In fact, the early Universe would become transparent for radiation even if
formation of hydrogen would have been forbidden, see e.g. \cite{Naselsky2011}.
The optical depth to Thomson scattering is%
\begin{align}
\tau &  =\int_{t}^{t_{0}}\sigma_{T}n_{b}cdt\simeq\nonumber\\
&  4\times10^{-2}\frac{\Omega_{b}}{\Omega_{m}}h\left\{  \left[  \Omega
_{\Lambda}+\Omega_{m}\left(  1+z\right)  ^{3}\right]  ^{1/2}-1\right\}  ,
\end{align}
where $\sigma_{T}$\ is the Thomson cross section, $\Omega_{i}=\rho_{i}%
/\rho_{c}$, $\rho_{c}=3H^{2}c^{2}/8\pi G$, $H=100h$ km s/Mpc and $b,m,\Lambda$
stand for, respectively baryons, dark matter and cosmological constant
contributions to the total energy density of the Universe. For large $z$\ we
have%
\[
\tau\left(  z_{\ast}\right)  =1,\qquad z_{\ast}\simeq8.4\Omega_{b}%
^{-2/3}\Omega_{m}^{1/3}h^{-2/3}.
\]
For typical values $\Omega_{b}h^{2}\simeq0.02$, $\Omega_{m}\simeq0.3$, and
$h$\ $\simeq0.7$ we have $z_{\ast}\simeq60$. At such redshift the Universe
would be expected to become transparent to Thomson scattering. That is exactly
what happens in plasma in GRB sources. Below we show that, unlike
radiation-dominated cosmological expansion where comoving quantities also
fulfill relations (\ref{acceleration}), the comoving temperature in GRB
outflows remains always high enough to prevent recombination of hydrogen.

\subsection{GRBs}

During both acceleration and coasting phases the comoving temperature
decreases with radius, see (\ref{acceleration}) and (\ref{coasting}). The
optical depth to Compton scattering may be computed and the corresponding
photospheric radius may be obtained, see \cite{RSV2011} where
ultrarelativistic outflows were analysed in details. In particular, comoving
temperature at the photosphere decreases with the baryonic loading $B$ both at
acceleration and coasting phases. However, when the outflow reaches the radius
$R_{s}=B^{-2}R_{0}$, the comoving temperature becomes independent from $B$. In
that regime the expression for the photospheric radius is%
\begin{equation}
R_{tr}=\left(  \dfrac{\sigma E_{0}B}{4\pi m_{p}c^{2}}\right)  ^{1/2},
\label{Rtr_photon_thin}%
\end{equation}
where $m_{p}$ and $\sigma$\ are proton mass and Thompson cross section,
respectively. Indeed, using (\ref{T0}), (\ref{Rtr_photon_thin}),
(\ref{acceleration}) and (\ref{coasting}) in the case of instant energy
release we have%
\begin{align}
T_{\min}  &  =BT_{0}\left(  \frac{R_{c}}{R_{tr}}\right)  ^{2/3}=
\label{TminS0}\\
&  =\left(  \dfrac{3}{4\pi a}\right)  ^{1/4}\left(  \dfrac{\sigma}{4\pi
m_{p}c^{2}}\right)  ^{-1/3}\left(  E_{0}R_{0}\right)  ^{-1/12}.
\end{align}
Notice how extremely insensitive this value is with respect to the remaining
parameters $E_{0}$ and $R_{0}$! Expressed in units of typical energy and size%
\begin{equation}
T_{\min}^{(s)}\simeq42\left(  E_{54}R_{8}\right)  ^{-1/12}\quad\mathrm{eV}.
\label{TminS}%
\end{equation}
In the case of gradual energy release with $\Delta t\gg R_{0}/c$ and constant
luminosity and mass ejection rate the initial temperature is%
\begin{equation}
T_{0}\simeq\left(  \dfrac{L}{16\pi\sigma_{SB}R_{0}^{2}}\right)  ^{1/4},
\end{equation}
and similar expression to (\ref{TminS0}) may be derived%
\begin{equation}
T_{\min}=\left(  \dfrac{1}{16\pi\sigma_{SB}}\right)  ^{1/4}\left(
\dfrac{\sigma}{4\pi m_{p}c^{2}}\right)  ^{-1/3}L^{-1/12}R_{0}^{1/6}\Delta
t^{-1/3}, \label{TminW0}%
\end{equation}
which may be rewritten, introducing $L=10^{50}L_{50}$ erg/s and $\Delta
t=1\Delta t_{1}$ s, as%
\begin{equation}
T_{\min}^{(w)}\simeq17L_{50}^{-1/12}R_{8}^{1/6}\Delta t_{1}^{-1/3}%
\quad\mathrm{eV}. \label{TminW}%
\end{equation}

Even if (\ref{TminW}) appears to be less stringent that (\ref{TminS}), they
are both quite insensitive to initial parameters. As a result, even if the
comoving temperature decreases very much compared to its initial value,
typically on the order of MeV, at the photospheric radius it is always well
above the characteristic temperature $0.3$ eV at which recombination happens
\cite{1990PhRvD..41..354B}, thus preventing formation of neutral hydrogen. In
fact, if such hydrogen would be formed the cross section of interaction of
expanding particles with the circumburst medium would drastically decrease. As
a consequence no aferglow would be observed.

A simplified way to look at this lower bound on the comoving temperature at
the photosphere is to say that if a fraction\ $\epsilon$ of solar mass is
released in the volume having radius $\delta$\ solar Schwarzschild radii, then
its minimum comoving temperature before transparency is%
\begin{equation}
T_{\min}^{(s)}\simeq66\left(  \epsilon\delta\right)  ^{-1/12}\quad\mathrm{eV},
\end{equation}
in the case of instant energy release and%
\begin{equation}
T_{\min}^{(w)}\simeq2.8\epsilon^{-1/12}\delta^{1/6}\Delta t_{1}^{-1/4}%
\quad\mathrm{eV},
\end{equation}
in the case of gradual energy release during time $\Delta t_{1}$. Clearly in
both cases $\delta>1$, and likely $\epsilon<1$. Notice, that while in the case
of instant energy release the lower bound on temperature decreases with
increasing $\delta$, it instead increases in the case of gradual energy release.

The baryon to photon ratio in GRB plasma like in cosmology is large. This
ratio may be estimated as%
\begin{equation}
\frac{n_{\gamma}}{n_{B}}=\frac{m_{p}}{m_{e}}\frac{1}{B\left\langle
\varepsilon\right\rangle }\simeq1.8\times10^{5}B_{-2}^{-1}\left\langle
\varepsilon\right\rangle ^{-1},
\end{equation}
where $B_{-2}=10^{-2}B$, and $\left\langle \varepsilon\right\rangle $ is
average photon energy in the source of GRB in units of electron rest mass
energy. Thus the optical depth of electrons is much larger than the one of
photons and it is given in \cite{deGroot1980}%
\begin{equation}
\frac{\tau_{e}}{\tau_{\gamma}}=\log\Lambda+\frac{n_{\gamma}}{n_{e}}\simeq
\frac{n_{\gamma}}{n_{B}},
\end{equation}
where $\Lambda$\ is the Coulomb logarithm. It means that electrons are kept in
equilibrium with photons when the latter already decoupled from them
\cite{1998MNRAS.300.1158G}. In other words, electrons are forced to keep the
local temperature of photons. This may lead to efficient Comptonization of the
photon flow when it is decoupled from plasma and is passing through electrons
having locally different temperature.

As soon as plasma gets collisionless, laboratory spectrum of photons, baryons
and electrons is maintained. If it was thermal at decoupling it will remain
so. This shows another difference with respect to cosmology, where energy of
all particles decoupled from the thermal bath decreases due to the
cosmological expansion. For that reason in cosmology only the shape of the
spectrum is conserved with expansion, but not the temperature.

Therefore, we have reached the conclusion that hydrogen recombination which is
responsible for transparency of cosmological plasma does not occur in GRB
plasma. This difference in physical conditions may result in deviations from
black body spectrum, as observed in GRBs.
Available studies of photospheric emission in GRBs in the literature show that
deviations from the perfectly thermal spectrum come mostly from three
effects:\ a) dynamical and ultrarelativistic character of plasma outflows and
geometric effects \cite{RSV2011}; b) "fuzzy photosphere" effect
\cite{2008ApJ...682..463P,2011ApJ...737...68B} and c) possible dissipation
mechanisms at the photosphere
\cite{2010MNRAS.407.1033B,2011MNRAS.415.1663T,2011MNRAS.415.3693R}. 

Recently we presented a theory of
photospheric emission from relativistic outflows, see \cite{RSV2011}. Assuming
that the spectrum of radiation in the comoving reference frame is the perfect
black body one, we have shown that the spectrum seen by a distant observer may
be essentially nonthermal due to both geometric and dynamical special
relativistic effects. The possibility that the spectrum of photospheric
emission is nonthermal also in the comoving frame is under investigation.

\section{Conclusions}

Regarding the dynamical aspects, there is an apparent similarity between the
electron-positron plasma in the early Universe and the one in GRB sources. For
an observer comoving with the radiation-dominated plasma in GRB source it may
look indistinguishable from a portion of radiation-dominated Universe.
However, this is true only in the absence of pressure gradients. Strong
gradients are likely present in GRB sources, and they should produce local
acceleration in the radiation-dominated electron-positron plasma, making it
distinct from the early Universe, where matter inhomogeneities are known to be weak.

There is also an apparent similarity with respect to the nucleosynthesis
phenomenon. Given that the temperature reached in GRB sources, see Eq.
(\ref{T0}), may be as high as several MeV, nuclear reactions are expected to
operate on timescales of $10^{-2}$ sec or shorter. That is on the order of
magnitude of dynamical timescale of the GRB sources. It means that
reprocessing of nuclear abundances may likely take place in GRB sources. Since
observations imply that GRBs may originate from compact stellar objects
elements heavier than helium are likely to be present in GRB sources. Such
heavy elements are then destroyed, resulting mainly in protons with small
admixture of helium. Thus, similarly to the early Universe, we do not expect
dynamically important free neutrons present in GRB plasma after it started to
expand and cool down unless they are engulfed by the expanding plasma later.

Finally, there is an important difference between the electron-positron plasma
in the early Universe and the one in GRB sources. We show in this paper that
unlike the primordial plasma which recombines to form neutral hydrogen, and
emits the Cosmic Microwave Background Radiation, GRB plasma does not cool down
enough to recombine.\ Therefore GRB plasma always becomes transparent due to
Compton scattering.


\begin{thebibliography}{10}

\bibitem{Ruffini2009}
R.~{Ruffini}, G.~{Vereshchagin}, and S.-S. {Xue},
\newblock Physics Reports {\bf 487}, 1 (2010).

\bibitem{2007PhRvL..99l5003A}
A.~G. {Aksenov}, R.~{Ruffini}, and G.~V. {Vereshchagin},
\newblock \prl {\bf 99}, 125003 (2007).

\bibitem{2008gcpa.book.....W}
S.~{Weinberg},
\newblock {\em {Cosmology}},
\newblock Oxford University Press, April 2008., 2008.

\bibitem{1993MNRAS.263..861P}
T.~{Piran}, A.~{Shemi}, and R.~{Narayan},
\newblock \mnras {\bf 263}, 861 (1993).

\bibitem{2000A&A...359..855R}
R.~{Ruffini}, J.~D. {Salmonson}, J.~R. {Wilson}, and S.-S. {Xue},
\newblock \aap {\bf 359}, 855 (2000).

\bibitem{1990ApJ...365L..55S}
A.~{Shemi} and T.~{Piran},
\newblock \apjl {\bf 365}, L55 (1990).

\bibitem{2010tsra.confE.274K}
E.~{Kafexhiu},
\newblock {Excitation and destruction of nuclei in hot astrophysical plasmas
  around black holes},
\newblock in {\em 25th Texas Symposium on Relativistic Astrophysics}, 2010.

\bibitem{2000NuPhS..80C0612D}
E.~V. {Derishev}, V.~V. {Kocharovsky}, and V.~V. {Kocharovsky},
\newblock Nuclear Physics B Proceedings Supplements {\bf 80}, C612+ (2000).

\bibitem{2010MNRAS.407.1033B}
A.~M. {Beloborodov},
\newblock \mnras {\bf 407}, 1033 (2010).

\bibitem{1968ZhETF..55..278Z}
Y.~B. {Zeldovich}, V.~G. {Kurt}, and R.~A. {Syunyaev},
\newblock Zhurnal Eksperimental noi i Teoreticheskoi Fiziki {\bf 55}, 278
  (1968).

\bibitem{1968ApJ...153....1P}
P.~J.~E. {Peebles},
\newblock \apj {\bf 153}, 1 (1968).

\bibitem{Naselsky2011}
P.~D. {Naselsky}, D.~I. {Novikov}, and I.~D. {Novikov},
\newblock {\em The Physics of the Cosmic Microwave Background},
\newblock Cambridge Astrophysics, Cambridge Univ. Press, 2011.

\bibitem{RSV2011}
R.~{Ruffini}, I.~A. {Siutsou}, and G.~V. {Vereshchagin},
\newblock arXiv:1110.0407  (2011).

\bibitem{1990PhRvD..41..354B}
J.~{Bernstein} and S.~{Dodelson},
\newblock \prd {\bf 41}, 354 (1990).

\bibitem{deGroot1980}
S.~R. {de Groot}, W.~A. {van Leeuwen}, and C.~G. {van Weert},
\newblock {\em Relativistic kinetic theory. Principles and Applications},
\newblock North Holland Publishing Company, 1980.

\bibitem{1998MNRAS.300.1158G}
O.~M. {Grimsrud} and I.~{Wasserman},
\newblock \mnras {\bf 300}, 1158 (1998).

\bibitem{2008ApJ...682..463P}
A.~{Pe'er},
\newblock \apj {\bf 682}, 463 (2008).

\bibitem{2011ApJ...737...68B}
A.~M. {Beloborodov},
\newblock \apj {\bf 737}, 68 (2011).

\bibitem{2011MNRAS.415.1663T}
K.~{Toma}, X.-F. {Wu}, and P.~{M{\'e}sz{\'a}ros},
\newblock \mnras {\bf 415}, 1663 (2011).

\bibitem{2011MNRAS.415.3693R}
F.~{Ryde} et~al.,
\newblock \mnras {\bf 415}, 3693 (2011).

\end{thebibliography}

\end{document}